\documentclass[pre,aps,twocolumn]{revtex4}
\usepackage{amsmath,amssymb,graphicx}

\begin{document}

\title{Nucleation times in the 2D Ising model}

\author{Kevin Brendel, G.T. Barkema and Henk van Beijeren}
\affiliation{Theoretical Physics, Utrecht University, Leuvenlaan 4,
3584 CE Utrecht, the Netherlands}

\date{\today}

\begin{abstract}
A theoretical framework is presented for the estimation of nucleation
times in systems with Brownian type dynamics. This framework is applied
to a prototype system: the two-dimensional Ising model with spin-flip
dynamics in an external magnetic field. Direct simulation results for
the nucleation times, spanning more than four orders of magnitude, are
compared with theoretical predictions. There are no free parameters; all
system properties required by the theory are determined by simulations
in equilibrium systems. Theoretical and simulation values are found to
agree in most cases within 20\%.
\end{abstract}

\maketitle

\section{Introduction}

Nucleation, the process of escape from a metastable state via the
spontaneous growth of a stable nucleus in a metastable surrounding,
has been studied often before, and excellent books and reviews exist
~\cite{nucleation}. A prototype system to study nucleation phenomena is
the well-known Ising model. Above the so-called critical temperature,
in absence of an external magnetic field, up- and down-pointing spins
are roughly equally abundant. Below the critical temperature, the
system prefers to be in either of two states: one state with a positive
magnetization in which most spins are pointing up, and one state with a
negative magnetization. In the presence of an external field one of these
states will become metastable, and will decay to the stable equilibrium
state with a certain nucleation rate.

In an earlier article~\cite{bbvb}, we presented a theoretical framework
to describe activated processes in systems with Brownian type dynamics. We
then tested this framework in the study of magnetization reversal times
in the two-dimensional Ising model. This first test-case was especially
simple, since many properties of the transition state, i.e. the state
with highest free energy on the trajectory from one free energy minimum
to the other, were well-known. For instance, it is a state in which half
the system has a positive, and the other half a negative magnetization,
and these two regions are separated by a pair of interfaces which span
the shorter dimension of the simulation cell.

Here, we apply this framework to nucleation phenomena. Starting in the
metastable state in which most spins are anti-aligned with the external
field, the system evolves via a transition state in which a droplet of
spins aligns with the field, to the stable minimum in which most spins
are aligned with the external field. This system is more complex than
that of magnetization reversal in the absence of an external field:
the size and shape of the critical droplet are not known a priori.
Acharyya and Stauffer studied nucleation times in the two- three- and
four-dimensional Ising model with external field, and found agreement
with predicitons of classical nucleation theory~\cite{acharyya98}. Here,
we study the two-dimensional case in much more detail.

The organization of our manuscript is as follows. In section~\ref{sec:model},
we describe the model that we study in detail. Next in
section~\ref{sec:theory}, we outline the theoretical framework, based on our
work on magnetization reversals, but extended to nucleation phenomena. We then
apply this framework to our prototypical model ---nucleation in the Ising model
with spin-flip dynamics. In section~\ref{sec:sim} we compare the theoretical
predictions with high-accuracy computational results.

\section{Detailed description of the model}
\label{sec:model}

We consider the Ising model on a square lattice with lateral dimension $L$,
periodic (helical) boundary conditions, and an external magnetic field $h$
which favors downward-pointing spins, with the Hamiltonian
\begin{equation}
H=-J\sum_{\langle i,j\rangle} s_i s_j + h\sum_i s_i,
\end{equation}
in which $s_i=\pm 1$ is the spin at site $i$, and $J$ is the coupling
constant. The summation runs over all pairs of nearest-neighbor sites;
those of site $i$ are $j=i\pm 1$ modulo $N$ and $j=i\pm L$ modulo $N$,
with $N= L^2$. The magnetization is defined as $M \equiv \sum_i s_i$;
it can take values $M=-N,-N+2,\dots,N$; all through this manuscript, we
restrict ourselves to systems in which $L$ is even. As a consequence,
$M$ takes only even values, and summations over a range of possible
magnetizations only run over even numbers, with an increment of 2.

The system evolves in time according to single-spin-flip dynamics with
Metropolis acceptance probabilities~\cite{metro}. If $C_i$ is the configuration
after $i$ proposed spin flips, a trial configuration $C'_{i+1}$ is generated by
flipping a single spin at a random site. This trial configuration is then
either accepted ($C_{i+1}=C'_{i+1}$) or rejected ($C_{i+1}=C_i$); the
acceptance probability is given by
\begin{equation}
P_a= \min \left[ 1, \exp(-\beta (E(C'_{i+1})-E(C_i))\right],
\end{equation}
in which $\beta=1/(k_BT)$ with Boltzmann constant $k_B$ and temperature $T$.
The time scale is set such that in one unit of time, on average each spin is
proposed to be flipped once. So in our system, in one unit of time we perform
$N$ Monte Carlo steps.

\section{Theoretical framework}
\label{sec:theory}

To study the behavior of nucleation times at temperatures below the
critical one we may consider an ensemble of a large number of systems
prepared in states with no large negative clusters present and study the
rate at which one of the clusters in the system grows beyond the critical
cluster size, after which the system is removed from the ensemble. The
spin-flip dynamics described above may be represented by a master equation
for the probability distribution $P({\bf S})$ of finding a system in the
state ${\bf S}$ at time $t$. Due to the huge number of possible states
this master equation cannot be solved analytically or even numerically
for system sizes of practical interest. Therefore, as an approximation
we assume that we may treat the clusters of negative spins in the system
as being independent. In this approximation, the probability that none
of the $N_c$ clusters has grown beyond the critical size, is simply the
$N_c^{th}$ power of the probability that a single cluster has not grown
beyond the critical size. The dynamics of a single cluster is then
modeled by a master equation for the equilibrium probability $P(C,t)$
that this cluster contains $C$ spins at time $t$:
\begin{eqnarray}
\frac{d P(C,t)}{d t} &=& \Gamma_{C,C+1}P(C+1,t)+\Gamma_{C,C-1}P(C-1,t)
\nonumber\\
&-& (\Gamma_{C+1,C}+\Gamma_{C-1,C})P(C,t),
\label{master}
\end{eqnarray}
with $\Gamma_{C',C}$ the transition rate from $C$ to $C'$. In section
\ref{sec:interdif} we describe how to estimate the transition rates
$\Gamma_{C+1,C}$.

In order that the equilibrium distribution be a stationary solution of
the master equation we impose the condition of detailed balance
\begin{equation}
\frac{\Gamma_{C+1,C}}{\Gamma_{C,C+1}}=\exp\left[ \beta(F(C)-F(C+1))\right],
\label{detbal}
\end{equation}
where, up to a constant, $\beta F(C)=-\ln P_{eq}(C)$, with $P_{eq}(C)$
the equilibrium probability of cluster size $C$.

The long-time nucleation rate as predicted by the master equation
(\ref{master}) follows as the largest eigenvalue (with a minus sign)
of this equation, supplemented with an absorbing boundary at $C=A$.
Here $A$ is an integer larger than the critical cluster size $C_x$,
chosen such that $P_{eq}(A) \gg P_{eq}(C_x)$, and clusters with size
$A$ are almost certain to nucleate. The absorbing boundary condition is
implemented by setting $\Gamma_{A-1,A}$ equal to zero.

The largest eigenvalue $-\nu$ of $\Gamma_{C,C'}$ in Eq.~(\ref{master}),
as well as the corresponding eigenvector $P_0(C)$, may be found by
requiring that the net current away from cluster size $C$ assumes the
value $\nu P_0(C)$. Using conservation of probability one easily checks
that this may be expressed as
\begin{eqnarray}
j_{C+1,C} &\equiv& \Gamma_{C+1,C}P_0(C) - \Gamma_{C,C+1}P_0(C+1) \nonumber \\
 &=& \nu\sum_{c\le C} P_0(c),
\label{eq:Gamma} \\
\nu &=& \frac{\Gamma_{A,A-1} P_0(A-1)}{\sum_{c\le A-1}P_0(c)},
\label{eq:nugamma}
\end{eqnarray}
where $j_{C+1,C}$ is defined as the net current flowing from $C$ to $C+1$.
This current may be approximated by
\begin{eqnarray}
j_{C+1,C} = \nu\frac{\sum_{c=1}^C \exp[-\beta F(c)]}
           {\sum_{c=1}^{C_x} \exp[-\beta F(c)]}& , &c\le C_x \nonumber \\
j_{C+1,C} = \nu &,& c\ge C_x \label{eq:current}
\end{eqnarray}
because the sum on the right hand side of Eq.~(\ref{eq:Gamma}) is dominated by
the terms with small $c$-values, for which $P_0(c)$ is approximately
proportional to $\exp[-\beta F(c)]$. This may be checked in hindsight against
the solution obtained. With this approximation the equation may be solved
recursively for $P_0(c)$ in terms of $P_0(A-1)$ for $c=A-2,\ A-3,\ \dots$, with
the result
\begin{equation}
\frac{P_0(c)}{P_0(A-1)}=\frac{\Gamma_{A,A-1}}{\nu} \sum_{m=c}^{A-1} 
      j_{m+1,m}\frac{\exp[\beta (F(m)-F(c))]}{\Gamma_{m+1,m}}.
\label{eq:mode}
\end{equation}
Since the sum over $m$ is dominated by values close to or larger than the
critical cluster size, for which $j_{m+1,m}$ is approximately equal to $\nu$,
we may replace Eq.~(\ref{eq:mode}) by
\begin{equation}
\frac{P_0(c)}{P_0(A-1)}=\Gamma_{A,A-1} \sum_{m=c}^{A-1} 
      \frac{\exp[\beta (F(m)-F(c))]}{\Gamma_{m+1,m}}.
\end{equation}
Now substituting this into Eq.~(\ref{eq:nugamma}) we arrive at the result
\begin{equation}
\nu = \left(\sum_{m=1}^{A-1} \frac{\exp\left[ \beta F(m)\right]}
	{\Gamma_{m+1,m}} \sum_{c=1}^{C_x}
	\exp\left[ -\beta F(c)\right]\right)^{-1},
\label{eq:nu}
\end{equation}
where we have used the fact that the sum over $c$ is dominated by small values
of $c$ to extend the sum over $m$ to $m=1$. The result in Eq.~(\ref{eq:nu}) is
well-known. It is usually derived by considering a state with a stationary
current in which mass is inserted at a constant rate on one side (e.g.\ at
$C=0$ in our case) and taken out as soon as it reaches the absorbing boundary
(see e.g.\ \cite{hanggi}, section IV E). In that case the replacement of
$j_{m+1,m}$ by a constant is exact.

One may give an even more accurate representation of the long time
behavior of $P(C,t)$ for arbitrary initial distributions $P(C,0)$ that are
concentrated near the origin, by writing it in the form
\begin{equation}
P(C,t)=k P_0(C) \exp[-\nu t]\ {\rm for}\ t\rightarrow \infty,
\label{eq:plong}
\end{equation}
where $k$ is a constant which represents the overlap between $P(C,0)$ and
$P_0(C)$. This gives for the probability $S(t)$ that the system has not yet
nucleated at time $t$
\begin{equation}
S(t)=\sum_{C=1}^{A-1} P(C,t) 
    = \exp[-\nu(t-t_d)]\ {\rm for}\ t\rightarrow\infty,
\end{equation}
where $t_d$ is called the delay time. The value of $t_d$ may be obtained from
the projection of the initial cluster size distribution $P(C,t)$ onto the most
slowly decaying eigenfunction of the master equation, (\ref{master}). This
eigenfunction was given (with a different normalization) in Eq.~(\ref{eq:mode}):
\begin{eqnarray}
  \psi^{{r}}(n) & \equiv & 
    \frac{\nu}{P_0(A-1)\Gamma_{A,A-1}} P_0(n) \nonumber \\
   & = & \sum_{m=n}^{A-1} j_{m+1,m}
  \frac{\exp{[\beta (F(m)- F(n))]}}{\Gamma_{m+1,m}}.  \label{psiright}
\end{eqnarray}
with $j_{m+1,m}$ given in Eq.~(\ref{eq:current}). As a consequence of the
condition of detailed balance, Eq.~(\ref{detbal}), the corresponding left
eigenvector is obtained by multiplying $\psi^r(n)$ by $\exp{[\beta F(n)]}$, or
\begin{equation} \label{left}
  \tilde{\psi}^{{l}}(n) = \sum_{m=n}^{A-1}
  j_{m+1,m}\frac{\exp{[\beta F(m)]}}{\Gamma_{m+1,m}}. 
\end{equation}
Notice that for small values of $n$ this is virtually independent of $n$,
because only the largest $m$-values give important contributions to the sum.
Using the proper normalization of the leading eigenfunction, and approximating
$P(C,0)$ by $\delta_{C,1}$, one now finds immediately that $t_d$ follows
from
\begin{eqnarray}
\label{expt0}
  \exp{[\nu t_{{d}}]} =
  \frac
  {\displaystyle
    \tilde{\psi}^{{l}}(1)
    \sum_{n=1}^{A-1} \left( \exp{[-\beta F(n)]} \tilde{\psi}^{{l}}(n) \right)
  }
  {\displaystyle
    \sum_{n=1}^{A-1} \left( \exp{[-\beta F(n)]} 
      \left( \tilde{\psi}^{{l}}(n) \right)^2 \right)
  }.
\end{eqnarray}
Now, by subtracting unity on both sides, dividing by $\nu$, approximating sums
over $m$ from $n$ to $A-1$ by sums from 1 to $A-1$, where appropriate, and using
Eq.~(\ref{eq:current}) to replace $j_{m+1,m}$ appearing in the left-hand sum of
the nominator, one obtains
\begin{equation} \label{t0}
  t_{{d}} = \frac{\displaystyle
   \sum_{n=1}^{A-1} \sum_{m=1}^{n-1} \left(
      \frac{
      W(n)}{\Gamma_{m+1,m}
      W(m)}
      \sum_{k=1}^{\min[m,C_x]}
      W(k) \right)
  \tilde{\psi}^{{l}}(n)} 
 {\displaystyle
    \sum_{n=1}^{A-1} \left(
    W(n) \tilde{\psi}^{{l}}(n)
      \right) },
\end{equation}
with $W(n)\equiv\exp(-\beta F(n))$.

\section{Simulations and results}
\label{sec:sim}

If one wants to apply the above theoretical framework to nucleation
times in the Ising model, the two ingredients required are: (i) the
equilibrium probability $P_{eq}(C)$ to find a cluster of size $C$;
and (ii) the transition rates $\Gamma_{C',C}$ from cluster size $C$ to
$C'$. We obtain these two ingredients via two different computational
approaches. In all our simulations, we use a technique known as multispin
coding~\cite{newman99}, which enables us to reach long simulation times
and thus good statistics.

\subsection{free energy landscape}
\label{sec:landscape}

We measure the distribution of cluster sizes for various values of $\beta$
and $h$, in a system with 64$\times$64 spins. We are actually only
interested in the distribution of clusters smaller than a certain size
$A$, which was discussed in section \ref{sec:theory}. Therefore we define
a cut-off cluster size $C_{max}$, which is typically chosen to be 300. We
modify our algorithm in the following way: starting with a configuration
$C_i$ we perform a fixed number $M$ of Monte Carlo steps, and measure
the sizes of all the clusters of down-spins in the system. If there is a
cluster with more than $C_{max}$ spins we reject the new configuration,
and choose $C_{i+1}=C_i$. Otherwise we accept the new configuration as
our $C_{i+1}$. In both cases we add the distribution of cluster sizes
in $C_{i+1}$ to the histogram. Finally we find the free energy $F(C)$ from

\begin{equation}
\beta F(C)=-\ln \frac{N(C)}{\sum_{C\leq C_{max}} N(C)},
\label{freeenergy}
\end{equation}
where $N(C)$ is the average number of clusters of size $C$ in the histogram.

If $C_{max}$ is chosen too large, the various clusters in the system
influence each other. In particular, excluded volume effects around
large clusters suppress bigger clusters in the metastable state more
than smaller ones. As a consequence the distribution of cluster sizes
depends on $C_{max}$. To determine whether a certain value of $C_{max}$
is allowed, we check that for small clusters the free energy curve
coincides with the same curve, obtained with a lower value for $C_{max}$.

\begin{figure}
\includegraphics[width=8cm]{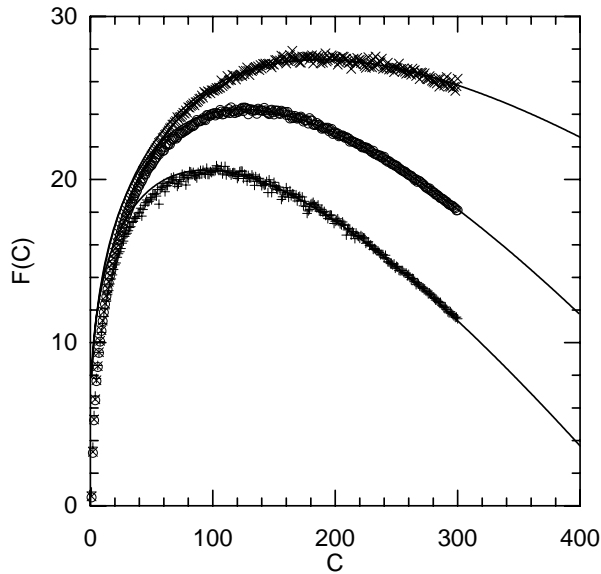} 
\caption{Free energy as a function of cluster size in the $64\times 64$ system
at $\beta J=0.58$ and $h=0.08$ ($\circ$), $\beta J=0.53$ and $h=0.08$ ($+$),
and $\beta J=0.56$ and $h=0.06$ ($\times$) . In the simulations, $C$ assumes
integer values $\ge 1$. The lines represent the Becker-D\"oring expression
Eq.~(\ref{bd}), with fitted values for $\sigma$.
\label{fig:fenergy}}
\end{figure}

Figure~\ref{fig:fenergy} shows our measurements for the free energy
according to Eq.~(\ref{freeenergy}) as a function of island size,
for some combinations of temperature and external field.

For large circular islands, classical nucleation theory can be used.
The free energy is then approximated by the Becker-D\"oring expression
\cite{becker}
\begin{equation}
F(C)\approx F_0+2 \sigma \sqrt{\pi C}-2hC,
\label{bd}
\end{equation}
where $F_0$ is some constant, $\sigma$ the excess free energy of the
interface per unit length, and $h$ the strength of the external field. We
fitted $F_0$ and $\sigma$ to the data in figure~\ref{fig:fenergy}, and
added the corresponding curves as lines in the same figure. As expected,
the measurements are well fitted by the curves, as long as the island
size is not too small.

\begin{figure}
\includegraphics[width=8cm]{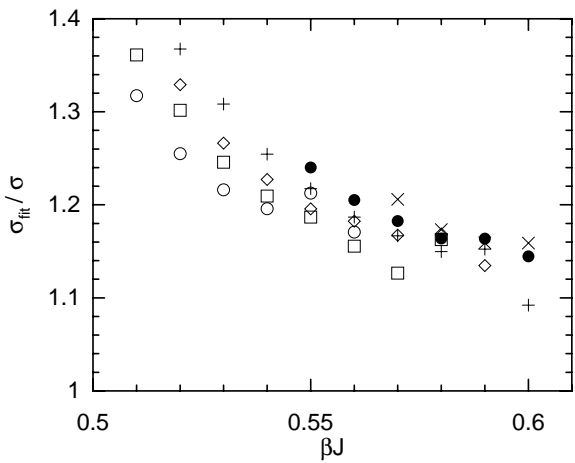} 
\caption{Surface tension obtained by fitting the free energy curve,
divided by the values given in Eq.~(\ref{eq:tension}), as a function of
$\beta J$. Different symbols denote different strengths of the external
field: $h=0.05 (\circ)$, $h=0.06 (\Box)$, $h=0.07 (\diamond)$, $h=0.08
(+)$, $h=0.09 (\bullet)$ and $h=0.10 (\times)$.
\label{fig:tension}}
\end{figure}

The surface tensions obtained in this way are larger than those given by the 
Onsager expression \cite{onsager}
\begin{equation}
\sigma=2J+\beta^{-1} \ln \tanh \beta J, \label{eq:tension}
\end{equation}
which is valid for long horizontal or vertical interfaces, and zero
external field (see figure \ref{fig:tension}). We will address this
issue in a separate paper~\cite{surftension}.

\subsection{interface diffusion coefficient} \label{sec:interdif}

\begin{figure}
\includegraphics[width=8cm]{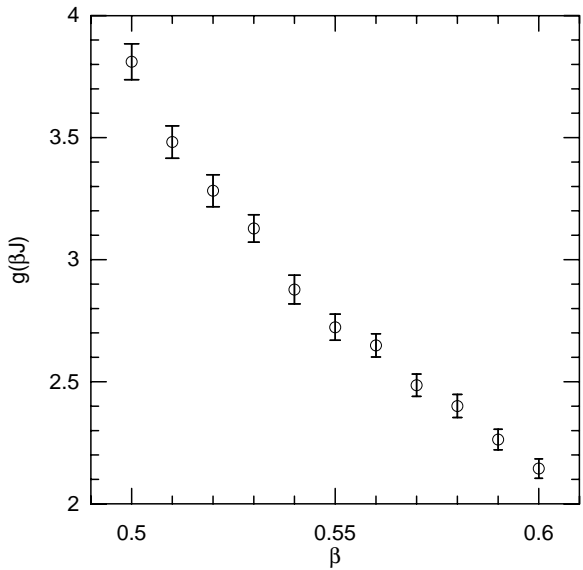}
\caption{Monte Carlo measurements of the diffusion coefficient per
interface length $g(\beta J)$, as a function of inverse temperature
$\beta J$.
\label{fig:g}}
\end{figure}

The second ingredient in the theoretical framework in
section~\ref{sec:theory} is the rate $\Gamma_{C+1,C}$ of cluster growth.
To estimate this rate we study the diffusion of a single interface in
a system with anti-periodic boundary conditions in the absence of an
external field, as described in \cite{bbvb}. The location of the interface
is obtained from the magnetization $M$. The diffusion coefficient $D$
is defined as:
\begin{equation}
D=\lim_{t\rightarrow \infty}
\left[\frac{\langle(M(t)-M(0))^2\rangle}{2t}\right]. \label{eq:D}
\end{equation}
This diffusion coefficient is, to a good approximation,
found to be linear in the length of the interface $B$:
\begin{equation}
D(B,L,\beta J)=g(\beta J) B+c, \label{eq:g}
\end{equation}
with the constant $c$ very close to zero.
The results for $g(\beta,J)$ for various temperatures are plotted
in figure \ref{fig:g}.

To arrive at an estimate for the rate $\Gamma_{C+1,C}$ for cluster growth
and shrinkage, we assume that the diffusion coefficient neither depends
on the external field nor on the shape of the interface (straight or
circular), but only on the length of the cluster boundary, which in
our case we set to $2\sqrt{\pi C}$. For large clusters this should be a
very good approximation. From Eqs.\ (\ref{eq:nu}) and (\ref{t0}) we see
that the contributions from small clusters to the expressions for $\nu$
and $t_d$ are almost insignificant, and inaccurate estimates of the
cluster boundary length for those clusters is not very important. This
then gives rise to a jump rate
\begin{equation}
\Gamma_{C+1,C}=\frac{g(\beta J)}{4}2\sqrt{\pi C}.
\label{jumprates}
\end{equation}
The factor 4 arises because the jumps in magnetization go by units
of 2. Notice that instead of identifying $\Gamma_{C+1,C}$ through $D$
we could have taken $\Gamma_{C,C+1}$ just as well. Due to the free energy
profile these quantities are not identical, though never much different,
but fortunately, for $C$-values close to the free energy maximum, which
are weighted most strongly, their difference becomes very small. In
addition there will be some compensating effects, because for $C\ge C_x$
one has $\Gamma_{C+1,C}>\Gamma_{C,C+1}$ whereas for $C<C_x$ this is just
the other way around.

\subsection{nucleation rates}
\label{sec:nuctimes}

To measure the long-time nucleation rates, we first bring the system into
equilibrium. At time $t=0$, we then reverse all the spins so that the
system is near its metastable equilibrium. We measure the time
the system needs to reach the magnetization corresponding to the stable
equilibrium. We make a histogram of all the measured times.

\begin{figure}
\includegraphics[width=8cm]{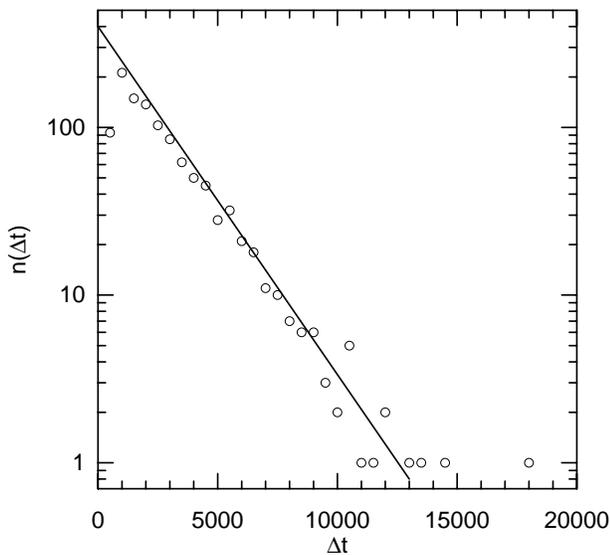}
\caption{Histogram of the time $\Delta t$ elapsed before nucleation
occurs, at inverse temperature $\beta J=0.54$ and external field strength
$h=0.08$, in a system with $64 \times 64$ sites. The straight line is
a fit, given by $n(\Delta t) \sim \exp(\Delta t/1990)$.
\label{fig:histo}}
\end{figure}

Figure~\ref{fig:histo} shows that at long times the decay function $f(t)$
behaves as $f(t)\sim\exp(-\nu_st)$. Here, we focus on the long-time behavior,
and are specifically interested in the escape rate $\nu_s$. We obtain this
quantity via a fitting procedure, in which we ignore the data up to a time
$t_0$, chosen such that $f(t)$ shows exponential time behavior for $t>t_0$.
Then we determine the time $t'$ at which half of the remaining events have
taken place. We then obtain the relaxation time $\tau \equiv \nu_s^{-1}$ from
$\tau=(t'-t_0)/\ln(2)$. Instead we could have made a linear fit of all the data
points beyond $t_0$, but this makes no significant difference. Table
\ref{tab:nuctime} gives the measured relaxation times $ \tau_1$. In the same
table, we also report the estimated relaxation times $ \tau_2=
(\left<N_c\right>\nu)^{-1}$ with $\nu$ as obtained in Eq.~(\ref{eq:nu}) and
$\left<N_c\right>\equiv \sum_{C=1}^A N(C)$ the average number of clusters in
the system. Finally we also give the ratio $\tau_2/\tau_1$. The table shows
that, while the relaxation times span more than four decades in time, the
estimated and measured relaxation times agree mostly within 20\%.

\begin{table}
\caption{
Measured values ($\tau_1$) and 
estimated values ($\tau_2$) for the
nucleation times in the $64\times 64$ system. \label{tab:nuctime}}
\begin{tabular}{lllll}
$\beta J$ \hspace{0.5cm} & $H$ \hspace{0.5cm} & $\tau_1$ \hspace{1.3cm} & 
$\tau_2$ \hspace{1.3cm} & $\tau_2/\tau_1$ \\
\hline
\hline
0.51 & 0.04 & $1.83\cdot 10^4$ & $2.26\cdot 10^4$ & 1.234 \\
0.51 & 0.05 & $3.35\cdot 10^3$ & $3.84\cdot 10^3$ & 1.146 \\
0.51 & 0.06 & $9.84\cdot 10^2$ & $1.14\cdot 10^3$ & 1.163 \\
0.52 & 0.04 & $1.07\cdot 10^5$ & $1.09\cdot 10^5$ & 1.019 \\
0.52 & 0.05 & $1.13\cdot 10^4$ & $1.31\cdot 10^4$ & 1.153 \\
0.52 & 0.06 & $2.84\cdot 10^3$ & $3.57\cdot 10^3$ & 1.257 \\
0.52 & 0.07 & $8.51\cdot 10^2$ & $1.09\cdot 10^3$ & 1.276 \\
0.52 & 0.08 & $3.96\cdot 10^2$ & $4.49\cdot 10^2$ & 1.134 \\
0.53 & 0.05 & $4.15\cdot 10^4$ & $4.79\cdot 10^4$ & 1.153 \\
0.53 & 0.06 & $8.94\cdot 10^3$ & $9.47\cdot 10^3$ & 1.059 \\
0.53 & 0.07 & $2.31\cdot 10^3$ & $2.57\cdot 10^3$ & 1.111 \\
0.53 & 0.08 & $8.98\cdot 10^2$ & $9.59\cdot 10^2$ & 1.068 \\
0.54 & 0.05 & $1.73\cdot 10^5$ & $2.13\cdot 10^5$ & 1.228 \\
0.54 & 0.06 & $2.90\cdot 10^4$ & $3.11\cdot 10^4$ & 1.071 \\
0.54 & 0.07 & $6.69\cdot 10^3$ & $7.27\cdot 10^3$ & 1.087 \\
0.54 & 0.08 & $2.17\cdot 10^3$ & $2.32\cdot 10^3$ & 1.068 \\
0.55 & 0.05 & $8.03\cdot 10^5$ & $1.13\cdot 10^6$ & 1.410 \\
0.55 & 0.06 & $9.57\cdot 10^4$ & $1.08\cdot 10^5$ & 1.126 \\
0.55 & 0.07 & $1.97\cdot 10^4$ & $2.08\cdot 10^4$ & 1.054 \\
0.55 & 0.08 & $5.60\cdot 10^3$ & $5.98\cdot 10^3$ & 1.070 \\
0.55 & 0.09 & $2.05\cdot 10^3$ & $2.17\cdot 10^3$ & 1.058 \\
0.56 & 0.05 & $4.02\cdot 10^6$ & $4.59\cdot 10^6$ & 1.140 \\
0.56 & 0.06 & $3.78\cdot 10^5$ & $4.17\cdot 10^5$ & 1.103 \\
0.56 & 0.07 & $5.81\cdot 10^4$ & $6.44\cdot 10^4$ & 1.109 \\
0.56 & 0.08 & $1.48\cdot 10^4$ & $1.57\cdot 10^4$ & 1.058 \\
0.56 & 0.09 & $4.92\cdot 10^3$ & $5.11\cdot 10^3$ & 1.038 \\
0.57 & 0.06 & $1.47\cdot 10^6$ & $1.36\cdot 10^6$ & 0.927 \\
0.57 & 0.07 & $1.84\cdot 10^5$ & $1.98\cdot 10^5$ & 1.080 \\
0.57 & 0.08 & $4.14\cdot 10^4$ & $4.40\cdot 10^4$ & 1.062 \\
0.57 & 0.09 & $1.24\cdot 10^4$ & $1.24\cdot 10^4$ & 1.000 \\
0.57 & 0.10 & $4.02\cdot 10^3$ & $4.53\cdot 10^3$ & 1.128 \\
0.58 & 0.06 & $5.80\cdot 10^6$ & $6.60\cdot 10^6$ & 1.139 \\
0.58 & 0.07 & $6.18\cdot 10^5$ & $5.72\cdot 10^5$ & 0.926 \\
0.58 & 0.08 & $1.10\cdot 10^5$ & $1.26\cdot 10^5$ & 1.143 \\
0.58 & 0.09 & $3.06\cdot 10^4$ & $3.11\cdot 10^4$ & 1.015 \\
0.58 & 0.10 & $1.13\cdot 10^4$ & $1.06\cdot 10^4$ & 0.938 \\
0.59 & 0.07 & $2.25\cdot 10^6$ & $2.48\cdot 10^6$ & 1.104 \\
0.59 & 0.08 & $3.49\cdot 10^5$ & $4.22\cdot 10^5$ & 1.209 \\
0.59 & 0.09 & $8.47\cdot 10^4$ & $8.36\cdot 10^4$ & 0.987 \\
0.59 & 0.10 & $2.31\cdot 10^4$ & $2.47\cdot 10^4$ & 1.068 \\
0.60 & 0.08 & $1.04\cdot 10^6$ & $1.19\cdot 10^6$ & 1.145 \\
0.60 & 0.09 & $2.05\cdot 10^5$ & $2.52\cdot 10^5$ & 1.229 \\
0.60 & 0.10 & $5.81\cdot 10^4$ & $6.72\cdot 10^4$ & 1.156 \\

\end{tabular}
\end{table}

\subsection{short-time behavior}
\label{sec:3D}

Besides the long-time exponential behavior of the nucleation probability, we
also study the deviation from this behavior at short times. To do this, the
master equation (\ref{master}) for the time evolution of one cluster is solved
numerically, with the initial condition that the cluster consists of a single
spin:
\begin{equation}
P(C,0)=\delta_{C,1}.
\end{equation}
We then compute the cumulative probability distribution $P_{nuc}(1,t)$
that the cluster has grown beyond a certain size $C_{max}$ during time
$t$. The corresponding probability distribution for the nucleation of one
of $N_c$ statistically independent clusters at $C_{max}$, at time $t$,
is given by
\begin{equation}
1-P_{nuc}(N_c,t)=\left( 1-P_{nuc}(1,t) \right)^{N_c}.
\label{nuc}
\end{equation}
The quantity $P_{nuc}(N_c,t)$ should be equal to the cumulative
distribution of nucleation times, if we use as a definition of nucleation
the first occurrence of a cluster of $C_{max}$ spins anywhere in the
system.

\begin{figure}
\includegraphics[width=8cm]{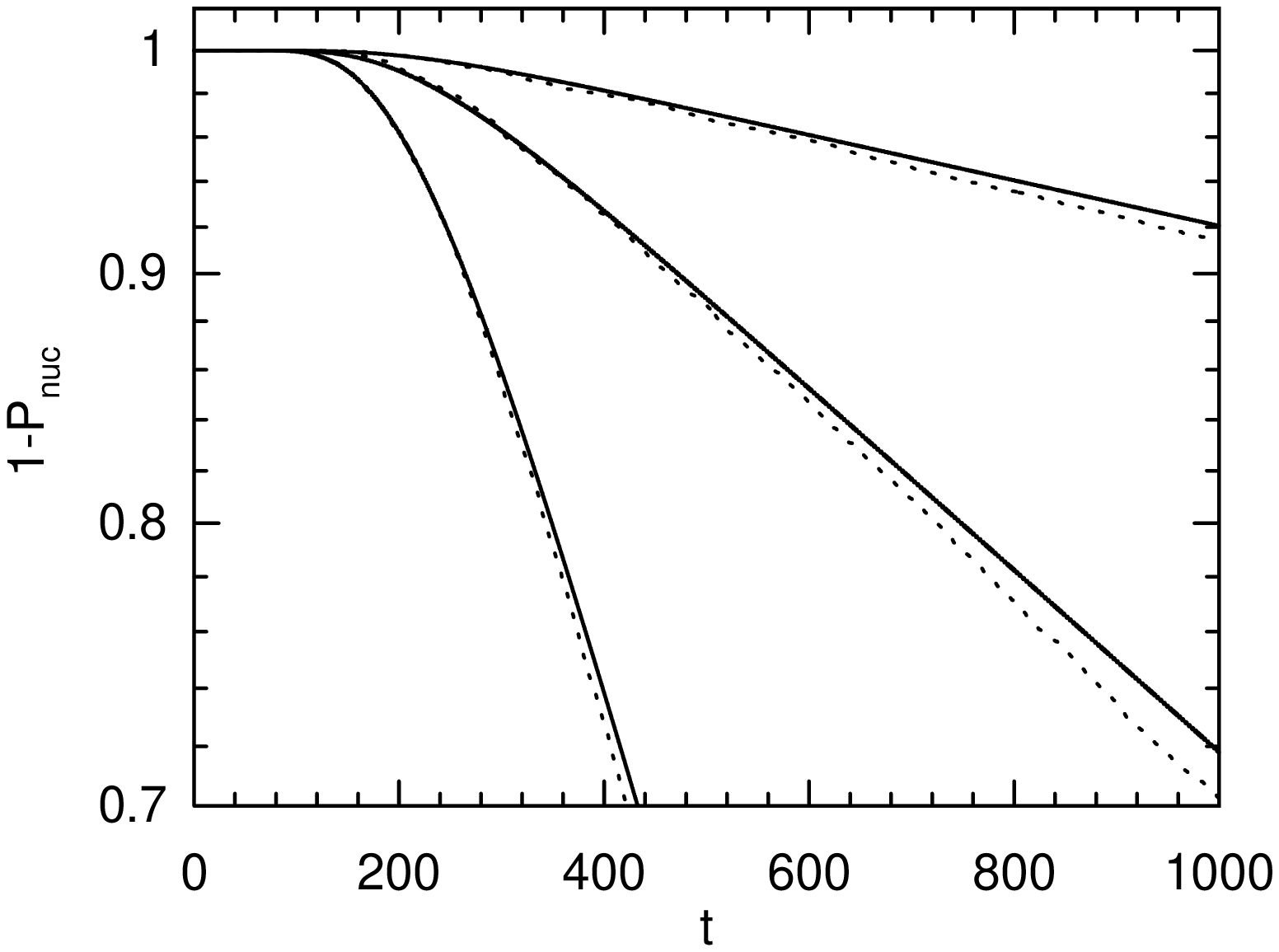}
\includegraphics[width=8cm]{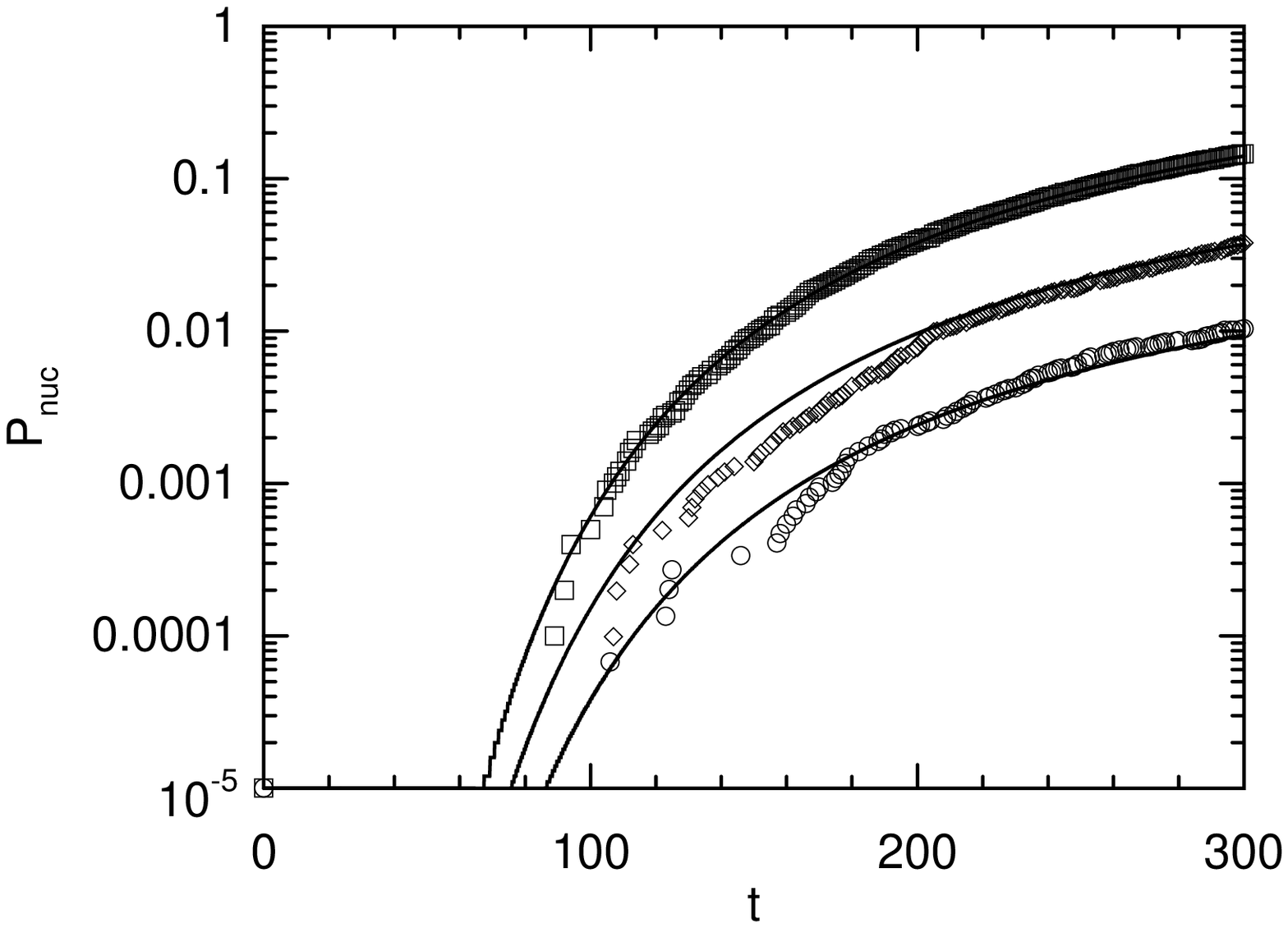}
\caption{$1-P_{nuc}(t)$ obtained from the time evolution of one cluster
(solid lines) according to the master equation (\ref{master}) combined
with Eq.~(\ref{nuc}), and by direct simulation of systems with $32\times
32$, $64\times 64$, and $128\times 128$ spins (the dotted lines) at
$\beta J=0.54$ and $h=0.08$. In both cases, the highest of the three
curves corresponds to the smallest system, and the lowest one to the
largest system. The lower plot shows the same data, but focuses on
short times.
\label{fig:short}}
\end{figure}

We have compared this result with the results of direct simulations, for
$\beta=0.54$ and $h=0.08$. We did this for three different system sizes:
$32\times 32$, $64\times 64$, and $128\times 128$. In all cases the starting
configuration was a system with zero magnetization. Within a few time steps
this develops into a quasi-equilibrium distribution as far as the distribution
of small cluster sizes is concerned, so for the present system sizes the fact
that we started from clusters of size zero rather than unity makes no
difference. Figure \ref{fig:short} shows the results. Here, we used for $N_c$
the mean number of clusters in the system as obtained from the same simulations
that were used to obtain the free energy as a function of cluster size.

The asymptotic slopes of the curves in the top panel of figure \ref{fig:short}
correspond to the nucleation rates. The times at which straight-line fits to
these curves cross $1-P_{nuc}=1$ correspond to the waiting times $t_d$. There
is excellent agreement between the direct simulations and the parameter-free
theoretical framework: The theoretical prediction obtained with Eq.(~\ref{t0})
is $t_d=234$ MC time units, while the fits vary between $t_d=228$ and 239 MC
time units. Also the behavior at short times is well described by the
theoretical framework, as is evidenced in the bottom panel of figure
\ref{fig:short}. The agreement at late times was discussed above.

\section{Discussion}
\label{sec:disc}

In this paper we showed that for two dimensional Ising models with spin-flip
dynamics Becker-D\"oring theory provides an excellent description of nucleation
time distributions, provided a realistic description is used for the free
energy of the growing droplets. We determined this free energy from cluster
size distributions in equilibrium Monte Carlo simulations and found it may be
fitted well by the Becker-D\"oring expression, provided one uses surface
tensions that are 10 to 20\% higher than the surface tension of a bulk
interface at zero field. Similar conclusions were reached by Auer and
Frenkel\cite{frenkel} in their studies of crystal nucleation in colloidal hard
sphere systems. They also determined cluster free energies by monitoring the
frequency of occurrence of clusters of a given size. Like we they could fit the
resulting curves quite well by a Becker-D\"oring expression with an effective
surface tension that exceeds the bulk surface tension by in their case 20 to
40\%.

A point one may question is whether the assumption of independent
noninteracting clusters holds even for clusters of small size. This may indeed
be doubted, but fortunately, at least for the asymptotic nucleation frequency
$\left<\nu N_c\right>$ this is not really relevant. One may choose to define as
clusters only those clusters that have a size $c \ge m_0$, with $m_0$ chosen
such that clusters of this size already are very rare, but still $\exp(\beta
F(m_0))\ll \exp(\beta F(C_x))$. From Eqs.\ (\ref{eq:nu}) and (\ref{freeenergy})
it then follows that \[
\nu_{m_o}=\frac{\sum_{c=1}^{C_x}N_c}{\sum_{c=m_0}^{C_x}N_c}\nu, \] since the
terms with $m<m_0$ in the first sum in (\ref{eq:nu}) basically do not
contribute. As a consequence of this the asymptotic nucleation rate in the
system is independent of the choice of $m_0$.

For the short-time nucleation rate the free energy of small clusters is also
important, especially if one starts from a state of zero magnetization.
However, if one starts from a quasi-equilibrium distribution representing the
state after a sudden reversal of the magnetic field, the short time behavior
will be dominated by clusters that were fairly large to start with and again
Eq.~(\ref{freeenergy}) may be trusted. In \cite{turtles} Van Beijeren gave
explicit expressions for the short-time behavior, which may be used if the
latter is well approximated by a diffusion equation in an external potential. 
In the present case these cannot be used, since on the relevant time scales the
hopping process between neighboring cluster sizes is not well-approximated by a
diffusion process. And since the hopping rates depend on cluster size no
analytic expressions for the short-time behavior are available. But the
numerical solution of the master equation (\ref{master}) gives a very good
agreement with the results from our Monte Carlo simulation of the nucleation
process, as was shown in Fig.\ \ref{fig:short}.

Besides the free energy as a function of cluster size our calculations require
the transition rates $\Gamma_{C,C\pm 1}$ between neighboring values of the
cluster size. These we estimated by setting them proportional to the mean
circumference of a cluster, determining the proportionality constant from the
simulated mobility of a straight interface in cylindrical geometry and imposing
the detailed balance condition (\ref{detbal}).

In our estimations we have been using a number of assumptions, whose validity
is not guaranteed under all conditions:

\begin{itemize}
\item Strong fields should modify the diffusion coefficient; this effect is
neglected. The freedom to modify the field strength within the metastable
region is limited though, and long nucleation times, as seen mostly in real
experiments, require weak fields.

\item The diffusion coefficient is assumed to be determined by the size of the
cluster alone, and is calculated on the assumption that this shape is strictly
spherical. This requires that the temperature is not too low, because at very
low temperatures the equilibrium shape of the cluster is squarish rather than
circular \cite{2deq}. On the other hand the temperature should not be too close
to the critical temperature for shape fluctuations to be reasonably small. To
some extent these fluctuations are taken into account, since our calculation of
the diffusion coefficient is done for a fluctuating interface around a
cylinder. But it is by no means certain that the fluctuations of a circular
interface are in all aspects comparable to those of the interface around the
cylinder.

\item It may happen that an island splits up or that two islands merge,
corresponding in our theoretical framework to non-zero transition rates
$\Gamma_{C,C+i}$ and $\Gamma_{C+i,C}$ with $i>1$. This effect also is partly
accounted for through the numerical determination of the diffusion constant on
a cylinder, but especially for larger islands the difference in geometry may
cause additional effects. However, these will only become important in systems
that contain a sizable density of islands of spins aligned with the external
field. Hence, also this approximation can be trusted least near the critical
point.

\item No memory effects are accounted for explicitly. For spin-flip dynamics
memory effects will chiefly be due to the influence of shape fluctuations on
the transition rates. Since shape fluctuations on larger length scales will
decay only slowly, these may be fairly long lasting effects. Again, our way of
determining the difusion coefficient will take many of these effects into
account implicitly, but memory effects involving large shape fluctuations may
be different for the present cluster geometry. For magnetization conserving
dynamics (consisting e.g. of local spin exchanges) much stronger memory
effects exist due e.g. to the effect that a spin that is released from a large
cluster has high probability of reattaching to it soon.

\end{itemize}

Under conditions in which the effects above are negligeable, our theoretical
framework is able to estimate nucleation rates with an accuracy in the range of
20\%. The small systematic overestimation by about 10\% of the nucleation time
by theory may have several causes. The radius of a cluster will be slightly
larger than our estimate because especially a large cluster will typically
contain a few holes in its interior. But in the investigated range of
temperatures this should not amount to a correction of more than 3\%. The
assumption that the diffusion coefficient is independent of the orientation of
the interface also may be not entirely correct. At low temperatures certainly a
diagonal interface is much more mobile than a straight one, but at the present
fairly elevated temperatures one would not expect this effect to be large.
Further there could be effects from the possibility of cluster splittings and
mergings, though some of these certainly are accounted for through the
numerical determination of the diffusion constant on a cylinder.

We are currently extending our investigations to two-dimensional Ising systems
with magnetization-conserving dynamics as well as to three-dimensional Ising
systems.


\begin{thebibliography}{*99}

\bibitem{nucleation}
P. G. Debenedetti, {\it ``Metastable Liquids''},
(Princeton University Press, Princeton, 1996).

\bibitem{bbvb}
K.\ Brendel, G.\ T.\ Barkema and H.\ van Beijeren, Phys.\ Rev.\ E {\bf 67},
026119 (2003). 

\bibitem{acharyya98}
M. Acharyya and D. Stauffer, European Physical Journal B {\bf 5}, 571 (1998).

\bibitem{metro}
N.\ Metropolis, A.~W.\ Rosenbluth, M.~N.\ Rosenbluth, A.~H.\ Teller, and
E.\ Teller, J.\ Chem.\ Phys.\ {\bf 21}, 1087 (1953).

\bibitem{hanggi}
P.\ H\"anggi, P.\ Talkner and M.\ Borkovec, Rev.\ Mod.\ Phys.\ {\bf 62}, 251
(1990).

\bibitem{newman99} M.\ E.\ J.\ Newman and G.\ T.\ Barkema, {\it ``Monte Carlo
Methods in Statistical Physics''}, Oxford University Press, Oxford, 1999.

\bibitem{becker}
R.\ Becker and D.\ D\"oring, Annalen der Physik {\bf 5}, 24 (1935).

\bibitem{onsager} L.\ Onsager Phys.\ Rev.\ {\bf 65}, 117 (1944).

\bibitem{surftension}
H. van Beijeren, G.T. Barkema and K. Brendel, to be published.

\bibitem{frenkel}
S.\ Auer and D.\ Frenkel, Nature {\bf 409}, 1020 (2001).

\bibitem{turtles}
H.\ van Beijeren, J.\ Stat.\ Phys.\ {\bf 110}, 1397 (2003).

\bibitem{2deq} J.\ E.\ Avron, H.\ van Beijeren, L.\ S.\ Schulman and R.\ K.\ 
P.\ Zia, J.\ Phys.\ {\bf A15}, L81 (1982).

\end{thebibliography}
\end{document}